\documentclass{elsart}
\usepackage{natbib}
\usepackage{graphicx}
\usepackage{amssymb}
\begin{document}
\begin{frontmatter}

 \title{The Characteristics of Valley Phase as Predictor of the Forthcoming Solar Cycle}
 \author{Baolin Tan$^{1,2}$}
 \ead{bltan@bao.ac.cn}
\address{$^{1}$CAS Key Laboratory of Solar Activity, National
Astronomical Observatories of Chinese Academy of Sciences, Datun Road A20, Chaoyang district,
Beijing 100101, China}
\address{$^{2}$School of Astronomy and Space Sciences, University of
Chinese Academy of Sciences, Beijing 100049, China}

\begin{abstract}

Is Solar Cycle 24 anomalous? How do we predict the main features of a forthcoming cycle? In order to reply such questions, this work partitions quantitatively each cycle into valley, ascend, peak, and descend phases, statistically investigate the correlations between valley phase and the forthcoming cycle. We find that the preceding valley phase may dominate and can be predictor of the forthcoming cycle: (1) The growth rate in ascend phase strongly negatively correlates to valley length and strongly positively correlates to cycle maximum. (2) The cycle maximum strongly negatively correlates to valley length, and strongly positively correlates to cycle minimum. (3) The cycle period strongly negatively correlates to the valley variation. Based on these correlations, we conclude that the solar cycle 24 is a relatively weak and long cycle which is obviously weaker than Cycle 23. The similarity analysis also presents the similar result. The Cycle 25 is also inferred possibly to be a weak cycle. These results can help us understanding the physical processes of solar cycles.

\end{abstract}

\end{frontmatter}

\section{Introduction}

About two centuries ago, solar cycles with period of about 11 years were discovered and then named as solar Schwabe cycle (Schwabe 1844). Traditionally, people numbered the cycle occurring during 1755 - 1766 as Cycle 1, and now it is in the Cycle 24. The solar cycle reflects intrinsically variations of solar activity, including the occurrence of solar flares, coronal mass ejections (CMEs), solar total irradiance, and the impact on the earth environment. Why is the present Cycle 24 so weak? Is it an unique or anomalous? What factors dominate the length, amplitude, and variations of a forthcoming cycles? Such questions are always puzzling and attracting the attentions of astrophysicists and publics widely. The nature of solar cycles remains one of the oldest and biggest unsolved problem in solar physics (Babcock 1961, Tan \& Cheng 2013, Charbonneau 2014, etc.).

Many indexes can be adopted to demonstrate solar cycles (Hathaway 2015, Usoskin 2017), such as numbers of solar flares, 10.7 cm radio flux (Holland \& Vaughn 1984), and solar magnetic field, galactic cosmic ray flux (Ferreira \& Potgieter 2004), and radioisotopes in tree rings or ice cores (Stuiver and Quay 1980, Solanki et al. 2004), etc. However, the most straightforward and longest records should be the sunspot number which had continuous data for more than 3 centuries. Since 2015 July, the original sunspot number data have been replaced by a new entirely revised series in WDC-SILSO, Royal Observatory of Belgium. This data series includes daily sunspot number since 1818, monthly mean sunspot number ($M_{1}$) and 13-month smoothed sunspot number ($M_{13}$) since 1749 to now. These data can be downloaded freely from website: http://www.sidc.be/silso/datafiles.

The periods and amplitudes are different from cycle to cycles. The cycle period varies from 9 year to 13.6 year, and the cycle's amplitude (maximum $M_{13}$) varies from $<$100 to $>$280. Many people attempt to predict the forthcoming cycles from different methods, such as statistical comparison (Du 2006), solar dynamo models (Choudhuri, Chatterjee and Jiang 2007, Jiang, Chatterjee and Choudhuri 2007, etc.), flux-transport dynamo (Dikpati, de Toma, and Gilman 2006), and other approaches (Jiang et al. 2015, Javaraiah 2015, Gopalswamy et al. 2016, Kakad, Kakad, and Ramesh 2017). However, so far, it is very difficult to reach a consensus in theories and practice.

Actually, a solar cycle is always featured with its minimum, ascend phase, maximum, and descend phase. In previous literatures, however, phase definitions are a bit vague. For example, Waldmeier (1935, 1939) found that the rise time was inversely proportional to the cycle amplitude, the so-called Waldmeier effect. Here, he defined the rise time as from minimum to the following maximum, and he neglected the fact that the growth rate varies significantly from minimum time to the maximum in a cycle. Actually, the growth rate is much slow near the minimum and maximum. Dikpati et al. (2008) found that Waldmeier effect might diminish in sunspot area data. The vague definition makes it difficult to obtain clear clues for understanding exact relationships between different phases and the real nature of solar cycles. It is necessary to partition each cycle into separated independent phases by an unified definition so that we can compare them qualitatively in different cycles.

This paper is organized as following: Section 2 proposed an unified qualitative definition to partition each solar cycle into valley, ascend, peak and descend phases. Section 3 presents statistical relationship between the valley phase and the main features of forthcoming cycles. Based on the above relationships, predictions of Cycle 24 and Cycle 25 are presented in Section 4. Finally, conclusions and discussion are summarized in Section 5.

\section{Definition and Partition of Phases of Solar Cycles}

This work select the well-known data of International monthly Sunspot Number ($M_{1}$, $M_{13}$) with 269 years continuous records (1749-2018) to investigate the phase characteristics of solar cycles. Fig. 1 presents a paragraph of profiles of $M_{1}$ and $M_{13}$ during Cycle 23 and 24 as an example which shows that even if around the cycle maximum, $M_{1}$ still fluctuates around the maximum value and lasts for a relatively long time. Similarly, the cycle minimum also lasts for a certain time. In order to investigate the statistic features and their relationships of solar cycles, it is necessary to define and partition each solar cycle into several independent phases quantitatively.

\begin{figure}[ht] 
\begin{center}
   \includegraphics[width=12 cm]{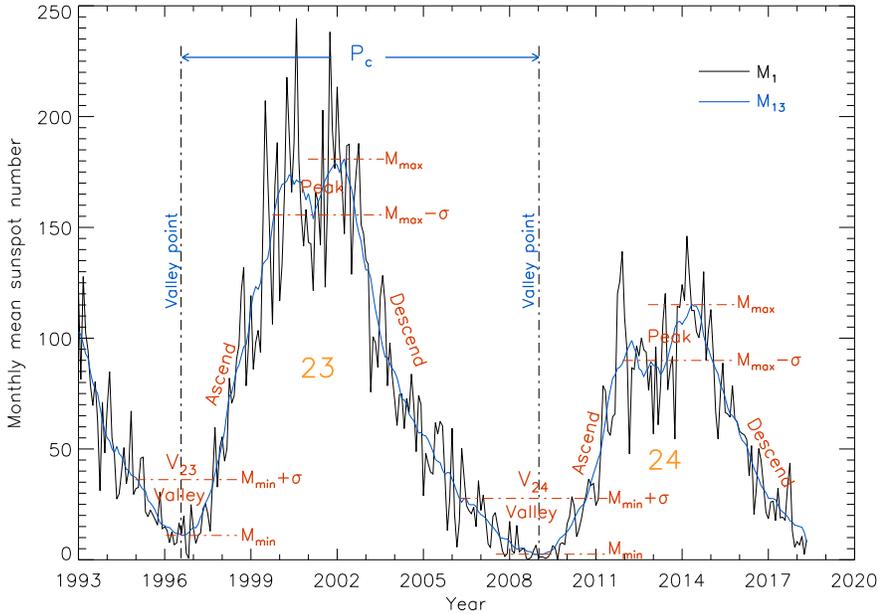}
\caption{Demonstration of phase partition of solar cycles. The black solid line is the recorded monthly mean sunspot number ($M_{1}$), and blue solid line is the 13-month smoothed $M_{1}$ ($M_{13}$). A solar cycle can be partitioned into four phases: valley, ascend, peak and descend phases. $\sigma$ is the averaged squared deviation between $M_{1}$ and $M_{13}$.}
\end{center}
\end{figure}

Because $M_{1}$ always fluctuates up and down around $M_{13}$, the magnitude of variation between $M_{1}$ and $M_{13}$ ($\sigma$) may define a certain duration around a certain $M_{13}$. Here, the variation is defined as:

\begin{equation}        
\sigma=\sqrt{\frac{\sum(M_{1i}-M_{13i})^{2}}{n}}.
\end{equation}

$n$ is the recorded number of $M_{1}$ and $M_{13}$. $\sigma$ will get different value in different paragraph of recorded data. When it is calculated with the total recorded data during 1749-2018, it is called total deviation, $\sigma=22.3$. When calculation is done in valley phase or in peak phase, it is called as valley deviation ($\sigma_{v}$), or peak deviation ($\sigma_{p}$), respectively. In this work, we use the total deviation $\sigma$ to define and partition phases of solar cycles uniformly and objectively.

(1) Valley phase, defined as a duration when $M_{13}\leq M_{min}+\sigma$ satisfied before a cycle maximum. $M_{min}$ is the minimum $M_{13}$ in a cycle. A valley phase is in the bottom between two cycles which has fewest sunspots. Its characteristics include cycle minimum $M_{min}$, valley length $L_{v}$, and valley deviation $\sigma_{v}$. The time of cycle minimum can be defined as a valley point (the vertical black dash-dotted lines in Fig. 1), two adjacent valley points define a solar cycle and its time length is defined as cycle period $P_{c}$. In valley phase, $M_{13}$ is decreasing slowly before valley point, then increasing slowly after the valley point.

A valley between cycle n-1 and cycle n is labeled as $V_{n}$. Therefore, valley phase is preceding the cycle maximum. For example, $V_{23}$ means the valley between cycle 22 and 23 which started from 1995 March, ended in 1997 September and lasted for 31 months with valley minimum of 11.3. In valley phase, the Sun is almost in a sleepy state with sporadic small solar eruptions. As it is preceding to a solar cycle, its characteristics may contain the precursor information of a forthcoming solar cycle.

(2) Ascend phase, defined as a period when $M_{13}$ rapidly increases after valley phase and before the following peak phase. The main feature of ascend phase includes its length $L_{a}$ and growth rate ($Gr_{a}$). The growth rate is defined as the increase of $M_{13}$ per month during the ascend phase. Here, the ascend length is different from the rise time defined by Waldmeier (1935, 1939), the later is a mixture of ascend phase and part of valley and peak phase.

(3) Peak phase, defined as the duration when $M_{13}\geq M_{max}-\sigma$ satisfied. $M_{max}$ is the maximum $M_{13}$ in a cycle. The main features of a peak phase include its maximum $M_{max}$, length $L_{p}$, and peak deviation $\sigma_{p}$. In peak phase, $M_{13}$ varies slowly, the Sun has the largest number of sunspots and the most frequently strong solar eruptions. As an example, the peak phase of Cycle 23 started from 1999 October, ended in 2002 July, and lasted for 34 months with cycle maximum of 179.1.

(4) Descend phase, defined as the period when $M_{13}$ gradually decreases after peak phase and before the next valley phase. The main feature of ascend phase includes its time length $L_{d}$ and decay rate ($Dr_{d}$). The decay rate is defined as the decrease of $M_{13}$ per month during the descend phase.

A solar cycle may span from one valley phase, through an ascend phase, a peak phase, and a descend phase, and finally falls into another valley phase. The period of a solar cycle ($P_{c}$) is between the two adjacent valley points (marked by the vertical black dash-dotted lines in Figure 1) of $M_{13}$. For example, Cycle 22 started from 1986 September and ended in 1996 July with period of 118 months. And Cycle 23 started from 1996 July, ended in 2008 December and lasted 149 month.

The advantage of a fixed $\sigma$ is that we can qualitatively compare them in different cycles. In the above partitions, we only use a fixed 1.0$\sigma$ to be the criterion to divide phases of solar cycles. However, we also test different criterions to partition the phases, for example, by fixed 0.7$\sigma$, 0.8$\sigma$, 0.9$\sigma$, 1.0$\sigma$, 1.1$\sigma$, 1.2$\sigma$ and 1.3$\sigma$. We found that 1.0$\sigma$ is the best criterion which can divide each solar cycle into valley, ascend, peak and descend separately, and each phase lasts for at least longer than 10 months. While a lower criterion may shorten the valley and peak phases, a higher criterion may shorted the ascend phase and even to make ascend phase vanished. Fig.2 and Fig.3 show profiles of valley and peak phases in the 24 solar cycles, respectively, partitioned by the criterion of 1.0$\sigma$.

\begin{table}\def~{\hphantom{0}}
  \begin{center}
  \caption{Phase Parameters of solar cycles since 1749. $L_{v}$, valley length (month). $L_{a}$, ascend length (month). $L_{p}$, peak length (month). $L_{d}$, descend length (month). $M_{min}$, cycle minimum. $M_{max}$, cycle maximum. $\sigma_{v}$, valley deviation. $\sigma_{p}$, peak deviation. $Gr_{a}$, growth rate (month$^{-1}$). $Dr_{d}$, decay rate (month$^{-1}$). $P_{c}$, cycle period (month).}
  \label{tab:kd}
  \begin{tabular}{lcccccccccccccc}\hline
     & $L_{v} $ & $M_{min}$ & $\sigma_{v}$ & $L_{a}$ & $Gr_{a}$ & $L_{p}$ & $M_{max}$ & $\sigma_{p}$ & $L_{d}$ & $Dr_{d}$  & $P_{c}$\\\hline
  1         &  41   & 14.0 & 12.0  & 45  & 1.85 & 13   & 141.7 & 24.8  & 38    & 2.07 & 135\\
  2         &  23   & 18.6 & 15.3  & 25  & 4.97 & 15   & 187.4 & 50.9  & 51    & 2.56 & 108\\
  3         &  23   & 12.0 & 11.9  & 22  & 9.36 & 10   & 262.9 & 51.2  & 55    & 3.68 & 112\\
  4         &  23   & 15.9 & 7.8   & 24  & 7.28 & 16   & 235.2 & 33.0  & 92    & 2.01 & 164\\
  5         &  50   & 5.3  & 9.0   & 12  & 2.62 & 52   & 81.4  & 12.9  & 17    & 2.16 & 146\\\hline
  6         &  75   & 0.0  & 6.1   & 24  & 1.59 & 26   & 82.7  & 27.7  & 35    & 1.09 & 152\\
  7         &  51   & 0.2  & 11.3  & 34  & 2.28 & 37   & 122.3 & 28.1  & 21    & 3.13 & 128\\
  8         &  25   & 12.2 & 9.5   & 21  & 8.93 & 12   & 244.0 & 53.3  & 59    & 3.08 & 116\\
  9         &  31   & 17.6 & 13.0  & 37  & 4.28 & 14   & 220.5 & 40.6  & 71    & 2.39 & 149\\
  10        &  27   & 6.0  & 6.8   & 24  & 5.68 & 22   & 187.0 & 24.8  & 64    & 2.06 & 135\\
  11        &  19   & 9.9  & 8.1   & 28  & 6.45 & 10   & 235.7 & 33.9  & 55    & 3.41 & 141\\
  12        &  50   & 3.7  & 8.4   & 40  & 1.86 & 17   & 123.0 & 26.8  & 25    & 2.78 & 136\\
  13        &  51   & 8.3  & 9.1   & 21  & 4.48 & 25   & 147.7 & 24.5  & 53    & 1.86 & 141\\
  14        &  48   & 4.5  & 7.0   & 17  & 3.27 & 52   & 104.5 & 34.1  & 20    & 2.86 & 140\\\hline
  15        &  48   & 2.5  & 8.9   & 31  & 4.03 & 11   & 172.2 & 43.2  & 47    & 2.53 & 118\\
  16        &  31   & 9.4  & 15.4  & 17  & 4.44 & 41   & 129.1 & 20.3  & 26    & 3.01 & 122\\
  17        &  39   & 5.8  & 8.5   & 24  & 6.91 & 20   & 197.1 & 38.5  & 50    & 2.78 & 125\\
  18        &  26   & 12.9 & 10.6  & 26  & 6.10 & 23   & 216.8 & 43.6  & 51    & 3.26 & 121\\
  19        &  24   & 5.1  & 8.6   & 26  & 9.10 & 15   & 286.9 & 39.9  & 63    & 3.62 & 126\\
  20        &  26   & 14.3 & 6.9   & 19  & 5.23 & 38   & 158.0 & 19.1  & 49    & 1.95 & 137\\
  21        &  32   & 17.8 & 13.0  & 23  & 7.36 & 18   & 231.7 & 25.2  & 47    & 3.69 & 126\\
  22        &  32   & 13.5 & 11.4  & 20  & 7.73 & 32   & 212.8 & 27.5  & 39    & 4.02 & 118\\
  23        &  30   & 11.3 & 7.0   & 25  & 4.94 & 34   & 179.1 & 30.0  & 50    & 2.64 & 149\\
  24        &  45   & 2.2  & 8.0   & 18  & 1.90 & 35   & 115.3 & 18.9  & --    &  --  & -- \\\hline
  Mean      & 36.3  & 9.3  & 9.7   & 25.1& 5.11 & 24.5 & 178.1 & 33.6  & 46.9  & 2.72 & 132.3 \\\hline
  \end{tabular}
 \end{center}
 \end{table}

\begin{figure*}[ht] 
\begin{center}
   \includegraphics[width=14 cm]{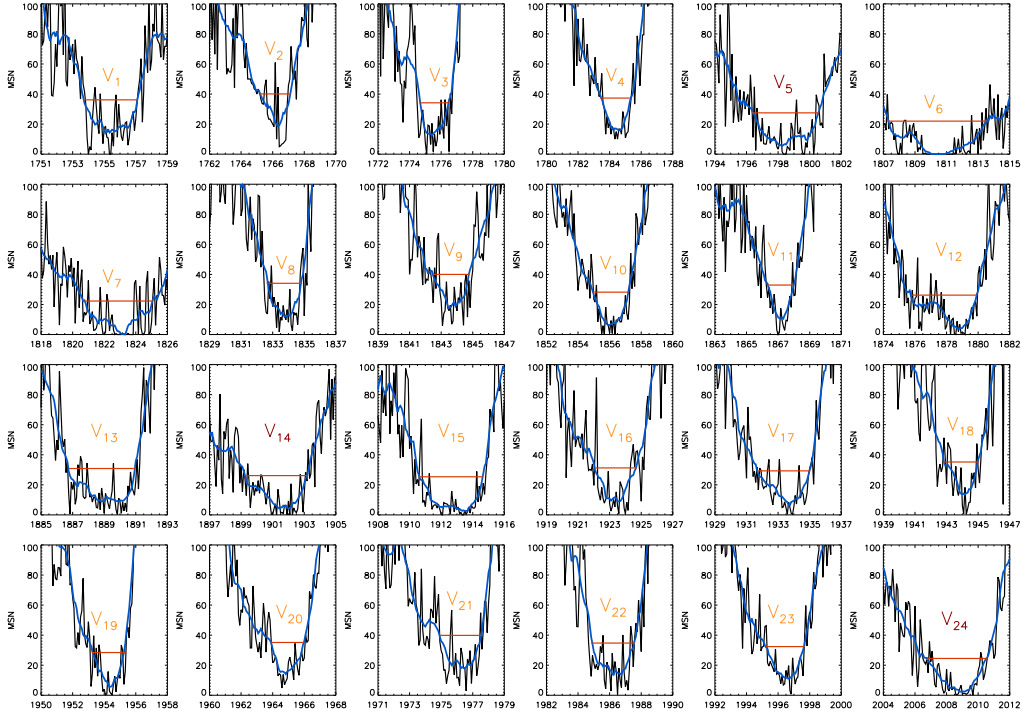}
\caption{Valley phases of the 24 solar cycles. The black solid line is the recorded $M_{1}$, and blue solid line is $M_{13}$. Red horizontal lines represent valley lengths.}
\end{center}
\end{figure*}

\begin{figure*}[ht] 
\begin{center}
   \includegraphics[width=14 cm]{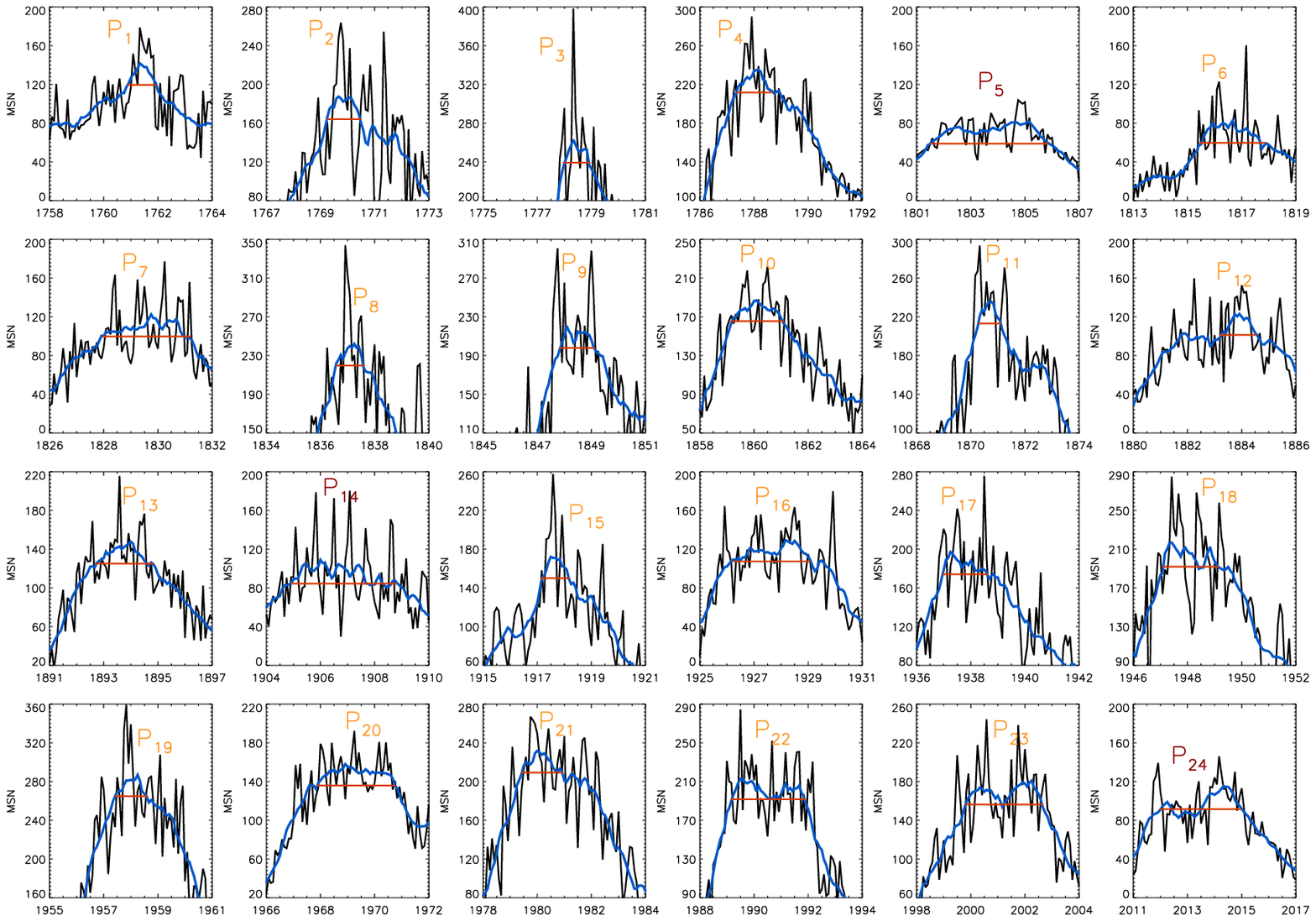}
\caption{Peak phases of the 24 solar cycles. The black solid line is the recorded $M_{1}$, and blue solid line is $M_{13}$. Red horizontal lines represent peak lengths.}
\end{center}
\end{figure*}

Table 1 lists statistic results of parameters in each phases among the 24 solar cycles since 1749. It shows that the averaged lengths of valley, ascend, peak, and descend phases are 36.3 months, 25.1 months, 24.5 months, and 46.9 months, respectively. The averaged peak deviation (33.6) is larger than 3 times of valley deviation (9.7). The averaged growth rate (5.11 month$^{-1}$) is about 2 times of decay rate (2.72 month$^{-1}$). Additionally, we find that the ascend phases of weak cycles are relatively short, while the strong cycles always have long ascend phases. This result is different from the Waldmeier rule. The possible reason is that our definition of ascend length is different from the rise time defined by Waldmeier (1935, 1939). The averaged cycle period is 132.3 months (11.03 years), and the longest one is Cycle 4 which lasted for 164 months (13.67 years) and was a relatively strong cycle with peak maximum $M_{max}$ up to 235.2 with the longest descend phase (92 months). The shortest one is Cycle 2, which lasted only 108 months (9 years) with maximum $M_{max}$ of 187.4, also a relatively strong cycle. Frick et al. (1997) found that the weaker the amplitude the longer the cycle. However, our statistics implies that the correlation between cycle length and the amplitude is not very strong (see Table 1 and the following Table 2 in Section 3). It is a bit of interesting that the longest solar cycle has only a relatively short peak phase (16 months), very close to the shortest peak phase (15 months). The strongest solar cycle is Cycle 19, which lasted for 126 months (10.5 years), and the maximum $M_{max}$ is 286.9 with the shortest single peak (15 months). The weakest solar cycle is Cycle 5, which lasted for 146 months (12.15 years) with maximum $M_{max}$ is 81.4, long valley (50 months) and the longest peak phase (52 months) with double peaks. It seems that valley length may anti-dependent to the cycle's amplitude.

\section{Relationships between the Characteristics of Solar Minima and the Main Features of Forthcoming Cycles}

The parameters of early phases, especially valley phases are much more meaningful for understanding and predicting the main features of the forthcoming cycles. Therefore, this work will mainly focus on investigation of the correlations between valley parameters and the forthcoming cycle's characteristics. Table 2 lists 55 correlation coefficients (Cor) between each pair of phase parameters among the 24 solar cycles.

\begin{table}\def~{\hphantom{0}}
  \begin{center}
  \caption{Correlation coefficients ($Cor$) between arbitrary two parameters of solar cycles. The means of parameters are same as in Table 1.}
  \label{tab:kd}
  \begin{tabular}{lcccccccccccc}\hline
& $L_{v} $ & $M_{min}$ & $\sigma_{v}$ & $L_{a}$ & $Gr_{a}$ & $L_{p}$ & $M_{max}$ & $\sigma_{p}$ & $L_{d}$ & $Dr_{d}$ \\\hline
  $M_{min}$  & -0.71 &       &         &       &        &       &       &         &       &        \\
$\sigma_{v}$ & -0.29 &  0.51 &         &       &        &       &       &         &       &        \\
  $L_{a}$    &  0.06 &  0.06 &   0.16  &       &        &       &       &         &       &        \\
$Gr_{a}$     & -0.76 &  0.49 &   0.12  & -0.27 &        &       &       &         &       &        \\
  $L_{p}$    &  0.35 & -0.28 &  -0.14  & -0.51 & -0.36  &       &       &         &       &        \\
  $M_{max}$  & -0.81 &  0.57 &   0.19  &  0.08 &  0.90  & -0.62 &       &         &       &        \\
$\sigma_{p}$ & -0.41 &  0.29 &   0.23  &  0.19 &  0.54  & -0.56 & 0.64  &         &       &        \\
 $L_{d}$     & -0.63 &  0.53 &  -0.09  &  0.08 &  0.62  & -0.64 & 0.76  &  0.44   &       &        \\
 $Dr_{d}$    & -0.50 &  0.20 &   0.39  & -0.06 &  0.60  & -0.14 & 0.58  &  0.34   & -0.05 &        \\
  $P_{c}$    & 0.27  & -0.07 &  -0.54  & -0.09 & -0.33  &  0.31 & -0.31 & -0.46   &  0.21 & -0.65  \\\hline
  \end{tabular}
 \end{center}
\end{table}

The main features of a forthcoming solar cycle includes cycle maximum ($M_{max}$), cycle period ($P_{c}$), growth rate in ascend phase ($Gr_{a}$), while the main characteristics may include the cycle minimum ($M_{min}$), valley length ($L_{v}$), and valley deviation ($\sigma_{v}$). From Table 1, it is puzzling that cycle period is uncorrelated to the cycle minimum ($Cor=-0.07$) and uncorrelated to valley length ($Cor=0.27$), cycle maximum is uncorrelated to the valley deviation ($Cor=0.19$), and growth rate is uncorrelated to the valley deviation, $Cor=0.12$ (Fig. 4). At the same time, it is only a slightly anti-correlation between cycle period and the maximum with correlation coefficients of -0.31. This result is consistent to Hathaway et al. (1994), Frick et al. (1997) and Solanki et al. (2002).

\begin{figure}[ht] 
\begin{center}
   \includegraphics[width=12 cm]{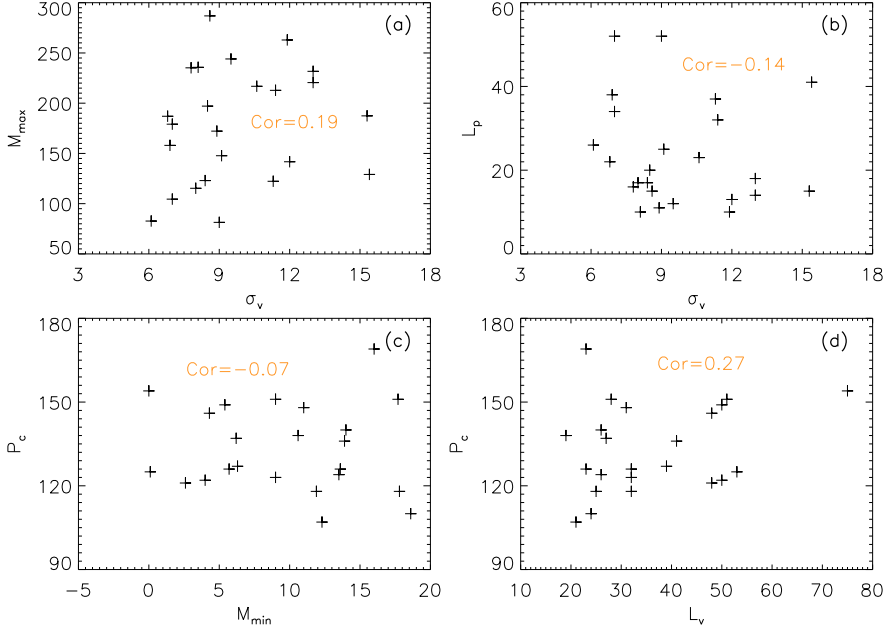}
\caption{Non-correlations between parameters of phases among 24 solar cycles. $L_{v}$, $\sigma_{v}$ and $M_{min}$ are valley length, valley deviation, and minimum, respectively. $M_{max}$, $L_{p}$ and $P_{c}$ are the maximum, peak length and the cycle period, respectively. The number in each panel is the correlation coefficient.}
\end{center}
\end{figure}

However, from Table 2 we find that there are several meaningful strong correlations between valley phase and the forthcoming solar cycle ($|Cor|\geq 0.54$, and the corresponding confidence level is above 99\%):

(1) The growth rate in ascend phase strongly negatively correlates to valley length ($Cor=-0.76$) and strongly positively correlates to the cycle maximum ($Cor=0.90$). We may use a function to fit the relation between growth rate in the forthcoming ascend phase and the preceding valley length (Fig.5a):

\begin{equation}        
G_{rA}\approx 11.2-0.18L_{v}\pm1.72.
\end{equation}

This fact indicates that long valley phase may predict a slow growth rate in the forthcoming ascend phases. For example, the cycle 6 has the longest valley phase (75 months) but with the slowest growth rate in the ascend phase (1.59 per month), while the cycle 3 has the fastest increasing ascend phase (9.36 per month) but with only 24 months valley phase. Our results are consistent with the conclusion obtained by Dikpati, Gilman and Kane (2010). They also found a strong anti-correlation between the length of a solar cycle minimum and depth of that minimum, a long minimum is both followed and preceded by weak cycles, and short minimum are followed and preceded by strong cycles. They explained this anti-correlation as due to the longer time available for annihilation of late cycle toroidal magnetic flux across the equator in the case of longer minimum.

(2) The cycle maximum strongly negatively correlates to valley length ($Cor=-0.81$), strongly positively correlates to cycle minimum ($Cor=0.57$). We can also use functions to fit these strong correlations (Fig. 5 b and c):

\begin{equation}        
M_{max}\approx 331.9-4.19L_{v}\pm35.8.
\end{equation}

\begin{equation}        
M_{max}\approx 85+8.65M_{min}\pm52.0.
\end{equation}

These results indicate that a long and weak valley phase may predict a relatively weak forthcoming cycle. This conclusion is different from the result obtained by Dikpati, Gilman and Kane (2010). The possible reason should be the different partition criterions. For example, Cycle 19 has maximum $M_{max}$ of 286.9 for its preceding valley phase is very short (24 month), while Cycle 5 has a very low maximum $M_{max}$ of 81.4 for its preceding very long valley phase (50 month). The strong correlation between the cycle minimum and maximum is consistent to the previous work (Hathaway et al. 1999), which means that the more the sunspot number during the valley phase, the strong the solar activity in the forthcoming cycle.

(3) The cycle period strongly negatively correlates to the preceding valley deviation ($Cor=-0.54$). This relation can be fitted by a function (Fig.5 d):

\begin{equation}        
P_{c}\approx 189.8-5.27\sigma_{v}\pm14.4.
\end{equation}

These strong correlations indicate that the perturbation during the preceding valley phase may strongly control the period of a forthcoming cycle. A high perturbation may predict a short forthcoming cycle. For example, Cycle 2 and Cycle 16 have the highest perturbation during their valley phases (15.3 and 15.4, respectively) and have the shortest cycles (108 months and 122 months, respectively), while the two longest cycles (164 months in cycle 4 and 152 months in cycle 6) only have small deviations in their preceding valley phase (7.8 and 6.1, respectively).

\begin{figure}[ht] 
\begin{center}
   \includegraphics[width=12 cm]{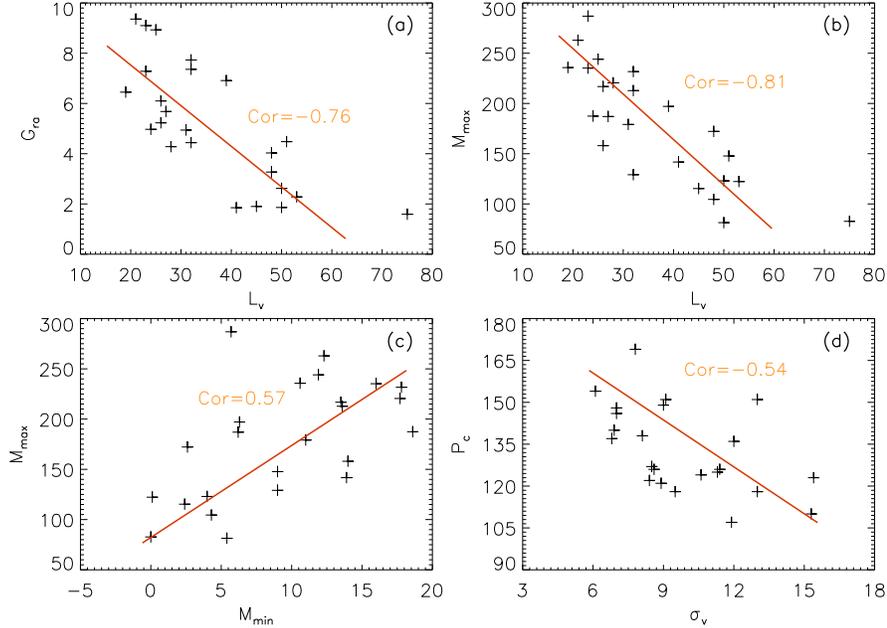}
\caption{Strong correlations between phase parameters of solar cycles. $L_{v}$, $\sigma_{v}$ and $M_{min}$ are valley length, valley deviation, and minimum, respectively. $M_{max}$ and $P_{c}$ are maximum and cycle period, respectively. Red solid lines are linear least square fitting results and the number in each panel is the correlation coefficient.}
\end{center}
\end{figure}

In summary, the cycle period is mainly dominated by the preceding valley deviation, the cycle maximum is mainly controlled by the minimum and the length of its preceding valley phase, and the growth rate is mainly dominated by the valley length. In a short, the behavior of the preceding valley phase may dominate the main features of a forthcoming solar cycle. Intrinsically speak, the variations of sunspot number reflect the change of solar magnetic fields and consequently the solar activities. The above evidences show that magnetic behaviors during the preceding valley phase may dominate the main properties of the forthcoming solar cycle. According to the theory of solar dynamo, the magnetic field during quiet Sun should be the seed fields for the next turn of dynamo processes.

\section{Predict Main Features of the Forthcoming Solar Cycles}

One of the important purposes of solar physics is to predict the forthcoming solar cycles. Many people attempted to predict the length of forthcoming solar cycles by using sunspot numbers near cycle minima (for example, Pishkalo, 2014, etc.). However, the relationship between cycle minimum and the forthcoming cycle period indicates that it seems no obvious correlation between these two parameters, for their correlation coefficient almost close to 0. From our statistic results, we find that it is the valley deviation ($\sigma_{v}$) which dominates the length of a forthcoming solar cycle, and the cycle minimum only affects on its maximum. At the same time, valley length also affects the cycle maximum and growth rate.

Here, we utilize the strong correlations obtained in Section 3 to study the main features of the forthcoming solar cycles. At first, let's consider the Cycle 24. Figure 1 shows that $V_{24}$ is started from 2006 November, ended in 2010 July, and lasted for 45 months with valley minimum of 2.2, valley deviation of 8.0. Based on Equation (3) and the valley length, we can predict the peak maximum of Cycle 24 is around 143.4$\pm$35.8. And from Equation (4) and valley minimum we may obtain another prediction of the peak maximum around 104.0$\pm$52.0. Both results are obviously lower than the maximum of Cycle 23 (179.1), and very close to the observed value (115.3). As for the period of Cycle 24, we may derive it from Equation (5) and the observed valley deviation, the calculated value is 152.9$\pm$14.4 months. It is very long and with a big uncertainty (about 9.5\%). Let's revisit Table 1. We find that the parametric characteristics of Cycle 24 is exactly similar to Cycle 5 and Cycle 14. They have similar long valley lengths (45-50 months), low valley minima (2.2-5.3), similar valley deviation (7.0-9.0) and low peak maxima (81.4-115.3). Fig. 6 presents the comparison of the profiles among these three weak cycles, which shows that their profiles are also very similar to each other. It is reasonable to assume that they will have similar cycle periods. As periods of Cycle 5 and 14 are 146 and 140 months respectively, it is reliable to predict the period of Cycle 24 will last at least 140 months or more.

\begin{figure}[ht] 
\begin{center}
   \includegraphics[width=10 cm]{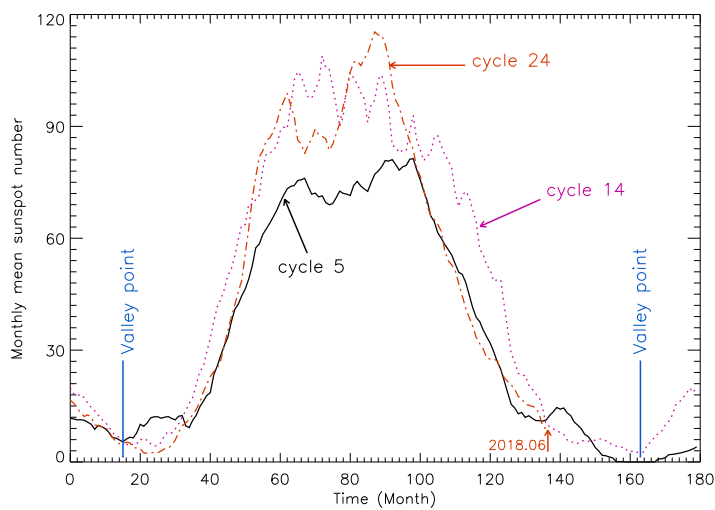}
\caption{Comparison of three weak cycles: Cycle 5, 14, and 24. Here the Cycle 24 is not complete, only contains the records before 2018 June.}
\end{center}
\end{figure}

Many people proposed that $V_{24}$ is an anomalously prolonged valley, which may imply that Cycle 24 will be peculiar weak and even means the advent of another Great Minimum (Nandy, Munoz-Jaramilo, Martens 2011, Broomhall 2017). Actually, this deduction seems to be lack of enough observational supports. The valley $V_{24}$ lasted for 45 months and the period of continuously no sunspot is only about one month (from 2008 July 21 to 2008 August 20, and from 2009 July 31 to 2009 August 31). $V_{24}$ is just composed of a simple monotonous decrease and a simple increase, while Table 1 shows that there are at least 7 prolonged valleys longer than $V_{24}$. Some of these prolonged valleys have long flat bottoms, for example, valley $V_{6}$  has a 20-month wide flat bottom with zero-$M_{1}$. Additionally, Cycle 24 is not the weakest cycle in the recorded history. Its minimum (2.2) is higher than that of $V_{6}$ and $V_{7}$, and its maximum is also higher than that of Cycle 5, 6 and 14. In a word, Cycle 24 is just an ordinary and relatively weak cycle.

Fig. 7 presents the details of Cycle 24. The above prediction of the period of Cycle 24 is at least 140 months, and it started from 2008 December, therefore, it will last to 2020 August or after. The yellow dashed line is postulated profile after 2018 June. Here, we find that the valley $V_{25}$ possibly started from about 2017 April. It will last to after 2022, and its valley length will exceed 56 months. From now on, the sunspot number will slowly and slowly decrease in several years. The minimum of $V_{25}$ will be very small. From Equation (3) and (4), we may conclude that the Cycle 25 will also be a relatively weak cycle. However, because the exact values of $V_{25}$ are not known at present, we can not predict the exact period and maximum of Cycle 25.

\begin{figure}[ht] 
\begin{center}
   \includegraphics[width=10 cm]{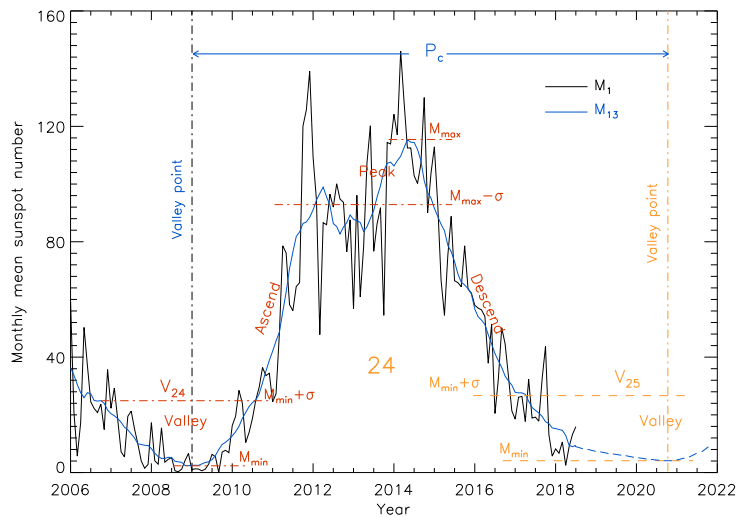}
\caption{Predicting results of solar cycle 24 and the main features of the valley phase of cycle 25. The yellow dashed line is postulated profile after 2018 June.}
\end{center}
\end{figure}

\section{Conclusions}

Based on the above phase analysis of solar cycles, we obtain the following conclusions:

(1) Based on the deviation between the monthly mean sunspot number and the 13-month averaged sunspot number recorded during 1749-2018, this work partitions each solar cycle into valley, ascend, peak, and descend phases qualitatively by an unified criterion. Each phase lasts for decades of months and changes from cycle to cycles. The advantage of unified fixed criterion of phase partition is that we can compare them qualitatively in different cycles.

(2) The statistic analysis shows that the preceding valley phase may dominate the main features of the forthcoming solar cycle, and can be predictor of the next solar cycle. Long valley phase may predict a slowly growth rate in the forthcoming ascend phase, low cycle maximum and a weak cycle. High cycle minimum may predict a high maximum and therefore a strong cycle. The valley deviation may dominate the period of a forthcoming cycle, for example, a big $\sigma_{v}$ may predict a short solar cycle and a small $\sigma_{v}$ will predict a long solar cycle.

(3) The above strong correlations may help us to predict the forthcoming solar cycles. Based on the strong correlations and the recorded data, we postulated that Cycle 24 is a relatively weak and long cycle, obviously weaker than Cycle 23. Its period will be at least 140 months, which will continue to after 2020 August. The similarity with historical recorded cycles also indirectly confirm the above conclusions. Additionally, we also predict that the $V_{25}$ will be a very long valley with a small minimum, which may imply that the Cycle 25 will also be a relatively weak cycle.

The correlations between valley phase and the forthcoming solar cycle indicate that valley phase may contain the important information of a forthcoming cycle, and this can help us to understand the physical processes of solar cycles. For example, the positive correlation between cycle maximum and the preceding valley minimum may imply that a relatively strong magnetic field in the valley phase will drive a strong solar cycle. The negative correlations between cycle period and the preceding valley deviation may imply that a complex valley phase may trigger a fast rising and decaying solar cycle which will have relatively higher maximum and short period, while a quiet valley phase may only drive a solar cycle with relatively lower maximum and longer period. However, so far we do not know the exact physical mechanism. It is meaningful to research the detailed physical mechanisms of relationships between the preceding valley phase and the forthcoming solar cycles. We will do it in the following studies.

\section*{Acknowledgments}

This work adopt the data of International Sunspot Number entirely revised by the team in WDC-SILSO, Royal Observatory of Belgium, and the research is supported by NSFC Grants 11433006, 11373039, 11573039, 11661161015 and 11790301.

\label{}

\end{document}